\documentclass[12pt]{article}
\usepackage{graphicx}
\usepackage{xcolor}
\usepackage{amsmath}
\usepackage{amssymb}
\usepackage{hyperref}
\usepackage{subfig}

\newcommand\pubdate{\today}

\def\Title#1{\begin{center} {\Large #1 } \end{center}}
\def\Author#1{\begin{center}{ \sc #1} \end{center}}
\def\Address#1{\begin{center}{ \it #1} \end{center}}

\newcommand\pubblock{\rightline{\begin{tabular}{l}  \\ 
         \pubdate  \end{tabular}}}
\newenvironment{Abstract}{\begin{quotation}  }{\end{quotation}}
\newenvironment{Presented}{\begin{quotation} \begin{center} 
             PRESENTED AT\end{center}\bigskip 
      \begin{center}\begin{large}}{\end{large}\end{center} \end{quotation}}

\begin{document}

\begin{titlepage}

\pubblock

\vfill
\Title{Low-$x$ physics at LHCb}
\vfill
\Author{Thomas Boettcher}
\Address{University of Cincinnati\\ on behalf of the LHCb collaboration}
\vfill
\begin{Abstract}
The LHCb detector's forward geometry provides unprecedented kinematic coverage
at low Bjorken-$x$. LHCb's excellent momentum resolution, vertex reconstruction,
and particle identification enable precision measurements at low transverse
momentum and high rapidity in proton-lead collisions, probing $x$ as small as
$10^{-6}$. In this contribution, we present recent studies of low-$x$ physics
using the LHCb detector. These studies include charged hadron, neutral pion, and
$D^0$ production in proton-lead collisions, as well as charmonium production in
ultraperipheral lead-lead collisions. Future prospects and implications for the
understanding of low-$x$ nuclear PDFs and parton saturation are also discussed.
\end{Abstract}
\vfill
\begin{Presented}
DIS2023: XXX International Workshop on Deep-Inelastic Scattering and
Related Subjects, \\
Michigan State University, USA, 27-31 March 2023 \\
     \includegraphics[width=9cm]{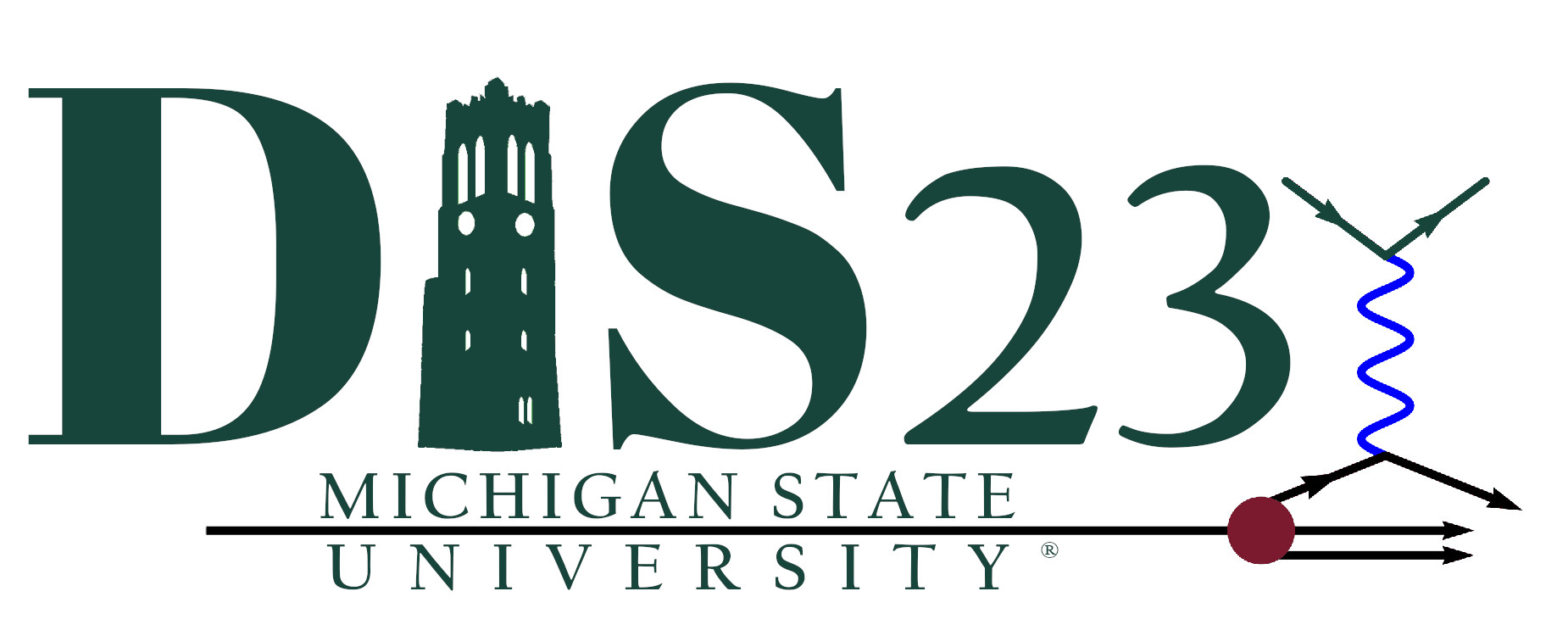}
\end{Presented}
\vfill
\end{titlepage}



\section{Introduction}

\begin{figure}[h!]
     \centering
     \subfloat[\label{fig:density}]{\includegraphics[width=0.44\textwidth]{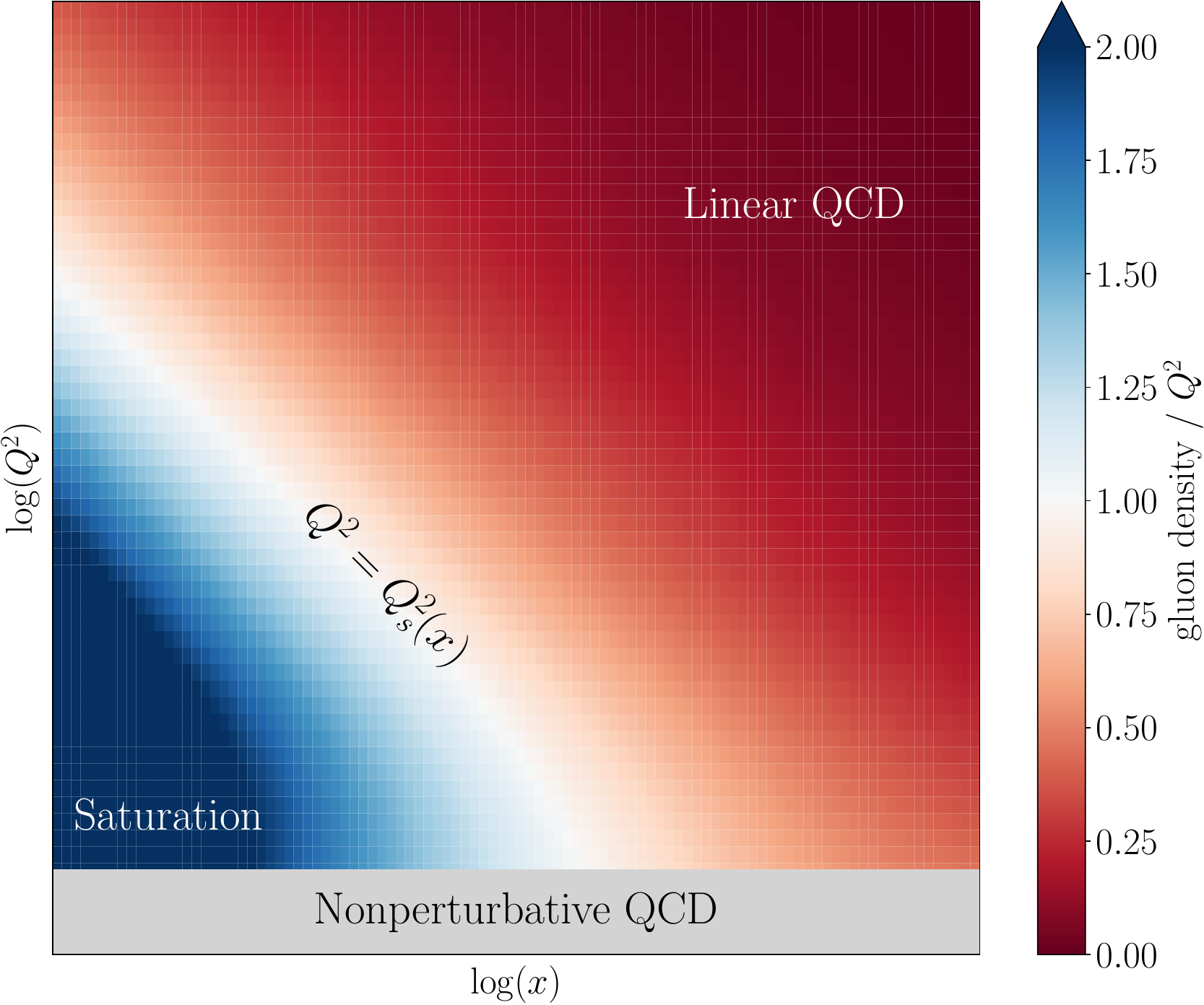}}~~
     \subfloat[\label{fig:coverage}]{\includegraphics[width=0.55\textwidth]{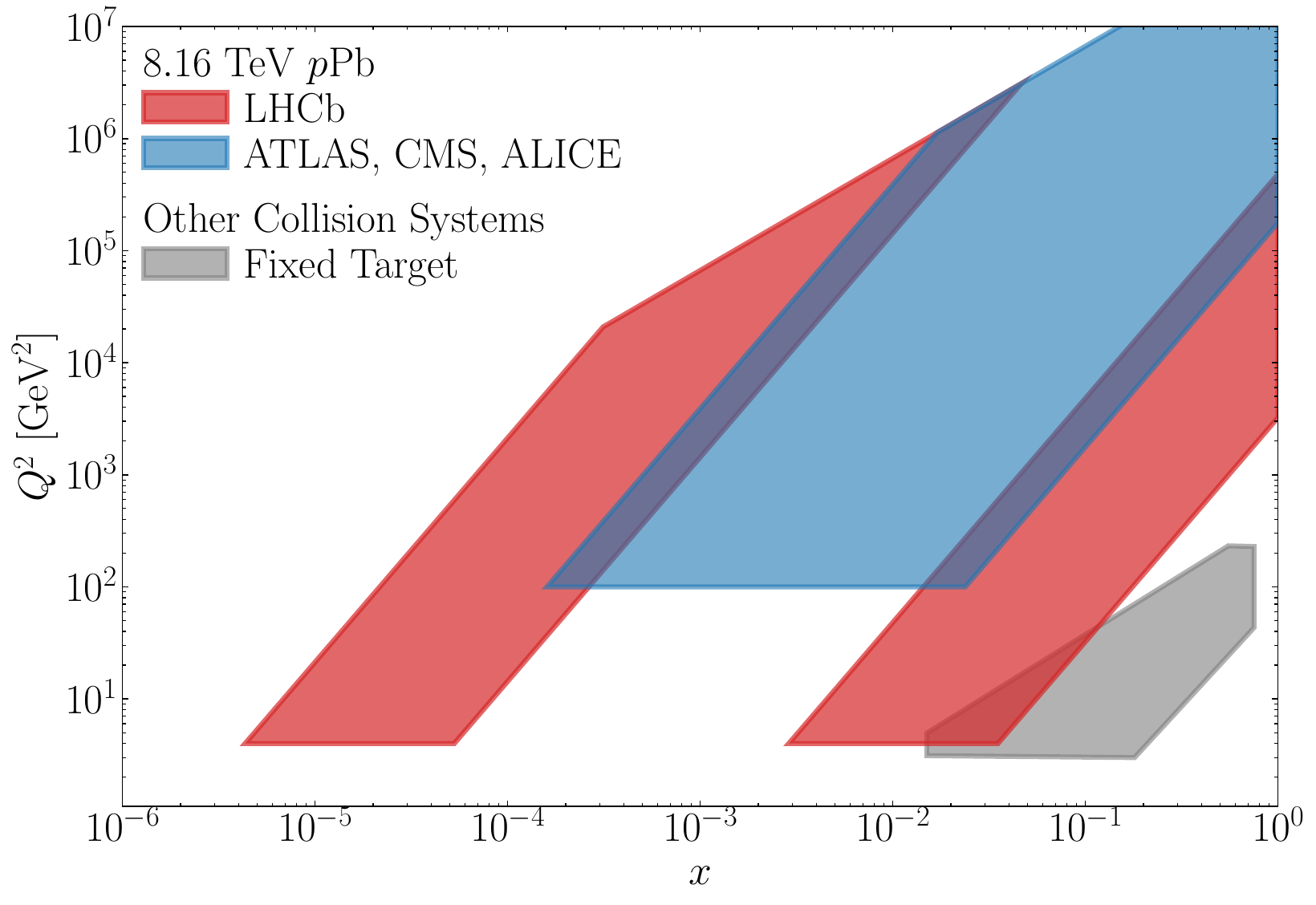}}
     \caption{(a) Illustration of the gluon area density in a nucleon divided by
     $Q^2$ as a function of $x$ and $Q^2$. For regions with values larger than
     one, saturation effects may be important. (b) The LHCb detector's coverage
     in the $x-Q^2$ plane compared to other LHC experiments and past
     fixed-target experiments.}
     \label{fig1}
\end{figure}

High-energy collisions at the LHC provide sensitivity to partons carrying small
fractions $x$ of the colliding nucleons' momenta. In this low-$x$ regime, parton
densities, and particularly the gluon density, in the nucleon are large. As a
result, interactions at a low momentum scale $Q$ can have a distance scale
larger than the typical spatial separation between gluons. This is illustrated
in Figure~\ref{fig:density}, which shows the gluon density in a nucleon divided
by $Q^2$. At large $x$ and large $Q^2$, partons are well-separated relative to
the length scale of the interaction. In this case, the evolution of parton
densities is dominated by QCD radiation and can be described using the linear
DGLAP equation. At small $x$ and small $Q^2$, parton separations become small
relative to the length scale of the interaction. This is particularly true in
heavy nuclei, as the gluon density scales with $A^{1/3}$. Consequently, the
study of nuclei at low $x$ is the study of matter at high gluon densities. In
this low-$x$ regime, which is often described using the color-glass condensate
(CGC) effective field theory~\cite{Kowalski:2007rw}, parton recombination
competes with radidiation, resulting in saturation of the parton density.

The LHCb detector is a fully instrumented detector in the forward region at the
LHC~\cite{LHCb:2014set}. Its forward acceptance of $2<\eta<5$ provides
sensitivity to very high- and low-$x$ partons. The LHCb detector features
excellent vertex resolution, charged-particle momentum resolution, and particle
identification capabilities, making it well-suited for studying the low-$p_{\rm
T}$ signals that are particularly sensitive to low-$x$ partons at low $Q^2$. The
LHCb detector collects $p{\rm Pb}$ data in two configurations: the forward
configuration, in which the proton travels toward the spectrometer at the time
of the collision; and the backward configuration in which the lead travels
toward the spectrometer. The forward configuration covers positive
pseudorapidity and is sensitive to low-$x$ partons in the nucleus, while the
backward configuration covers negative pseudorapidity and is sensitive to
moderate-to-high $x$. The LHCb detector's kinematic coverage in the $x-Q^2$
plane is shown in Figure~\ref{fig:coverage}.

\begin{figure}[h!]
     \centering
     \includegraphics[width=0.6\textwidth]{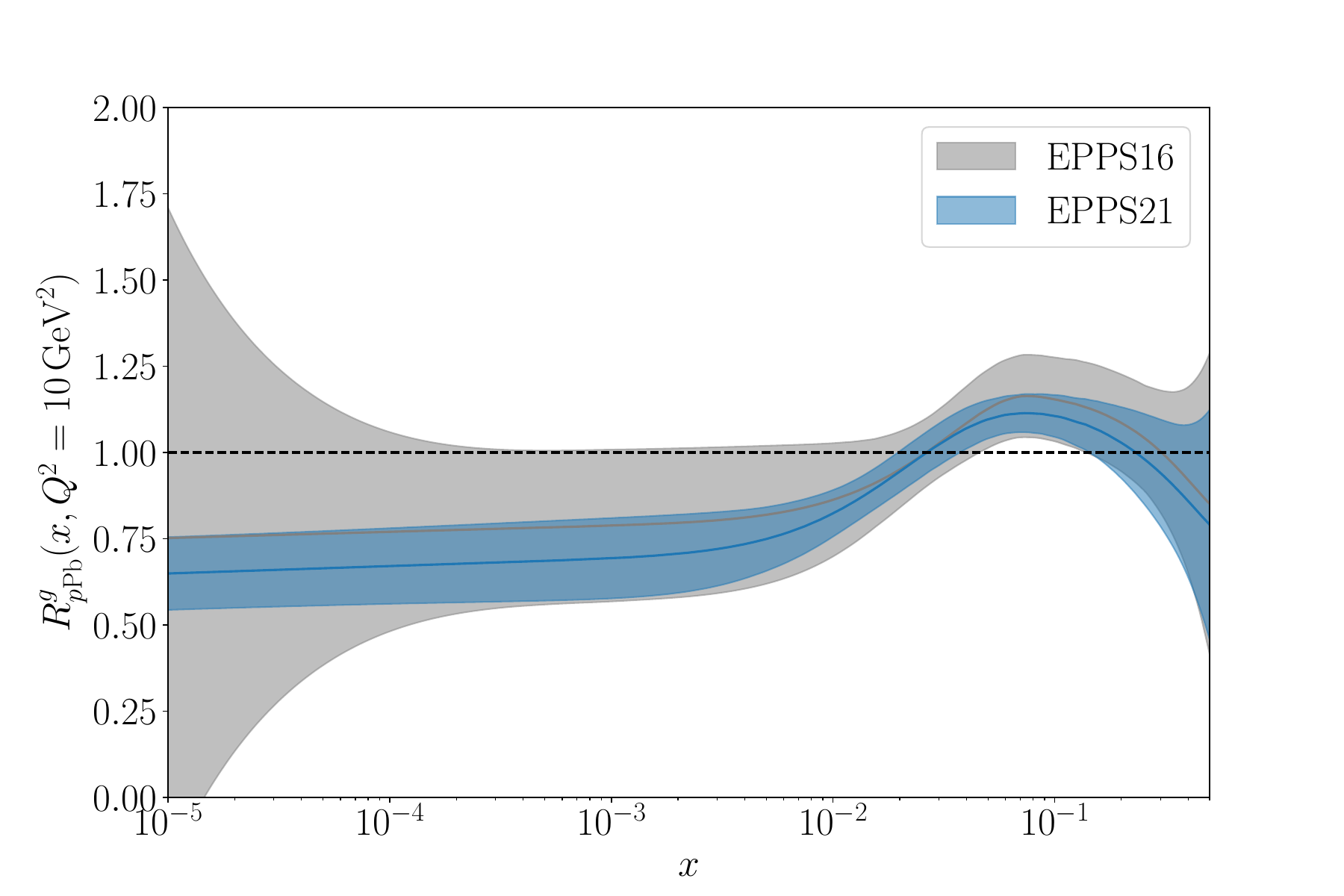}
     \caption{Comparison of the EPPS16\,\cite{Eskola:2016oht} and
     EPPS21\,\cite{Eskola:2021nhw} gluon nuclear modification factors. The
     EPPS21 fit uses LHCb $D^0$ production data, while the EPPS16 fit does not.}
     \label{fig:epps}
\end{figure}

Parton densities in hadrons are often described using parton distribution
functions (PDFs), and the parton densities of bound nucleons are described by
nuclear PDFs
(nPDFs)~\cite{Kovarik:2019xvh,AbdulKhalek:2022fyi,Eskola:2021nhw,Kovarik:2015cma,deFlorian:2011fp}.
PDFs and nPDFs encode the $x$ distributions of partons at a particular $Q^2$,
and their evolution in $Q^2$ is assumed to be governed by the linear DGLAP
equation. Because they encode the nonperturbative structure of protons and
nuclei, PDFs and nPDFs must be determined from data. Until recently, the gluon
nPDF was almost completely unconstrained for $x\lesssim 10^{-4}$. However,
recent LHCb measurements of open charm production in $p{\rm Pb}$ collisions at
$\sqrt{s_{\rm NN}}=5.02\,{\rm TeV}$\,\cite{LHCb:2017yua} have resulted in large
reductions in uncertainties in the gluon nPDF at low $x$ in state-of-the-art
nPDF fits~\cite{AbdulKhalek:2022fyi,Eskola:2021nhw}. The effect of these data is
shown in Figure~\ref{fig:epps}, which shows the ratio $R^g_{\rm Pb}$ of the
gluon distribution in a bound nucleon to the gluon distribution in the free
proton in fits with and without LHCb data. The LHCb data tightly constrain the
gluon nPDF at $x$ as low as about $10^{-5}$. Consequently, additional
measurements will serve to overconstrain nPDFs, potentially revealing the
effects of nonlinear QCD at low $x$.

\section{Light hadron production}

\begin{figure}[h!]
     \centering
     \includegraphics[width=\textwidth]{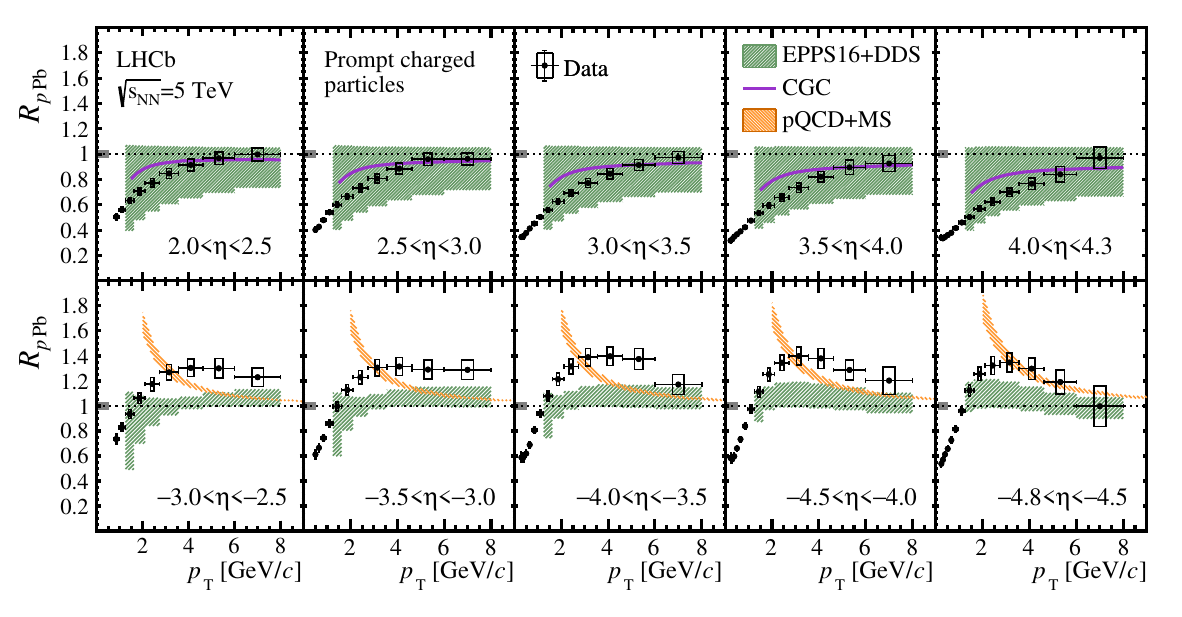}
     \caption{The charged-particle nuclear modification factor measured by LHCb
     in $p{\rm Pb}$ collisions at $\sqrt{s_{\rm NN}}=5.02\,{\rm
     TeV}$~\cite{LHCb:2021vww}.}
     \label{fig:chg}
\end{figure}

The minimum $Q^2$ of open charm production is limited by the relatively large
$D^0$ mass. In order to study lower $x$ and $Q^2$, the LHCb collaboration has
measured inclusive charged-particle production in $p{\rm Pb}$ collisions at
$\sqrt{s_{\rm NN}}=5.02\,{\rm TeV}$~\cite{LHCb:2021vww}. The measured nuclear
modification factor is shown in Figure~\ref{fig:chg}. At forward rapditiy, the
results agree with calculations using nPDFs, although the nPDF uncertainties are
large. The forward results show some tension with CGC calculations, but this
tension disappears when comparing to more recent next-to-leading order (NLO)
calculations~\cite{Shi:2021hwx}. The backward data, however, show a large
enhancement of charged particle production and are not successfully described by
any available models.

\begin{figure}
     \centering
     \includegraphics[width=0.8\textwidth]{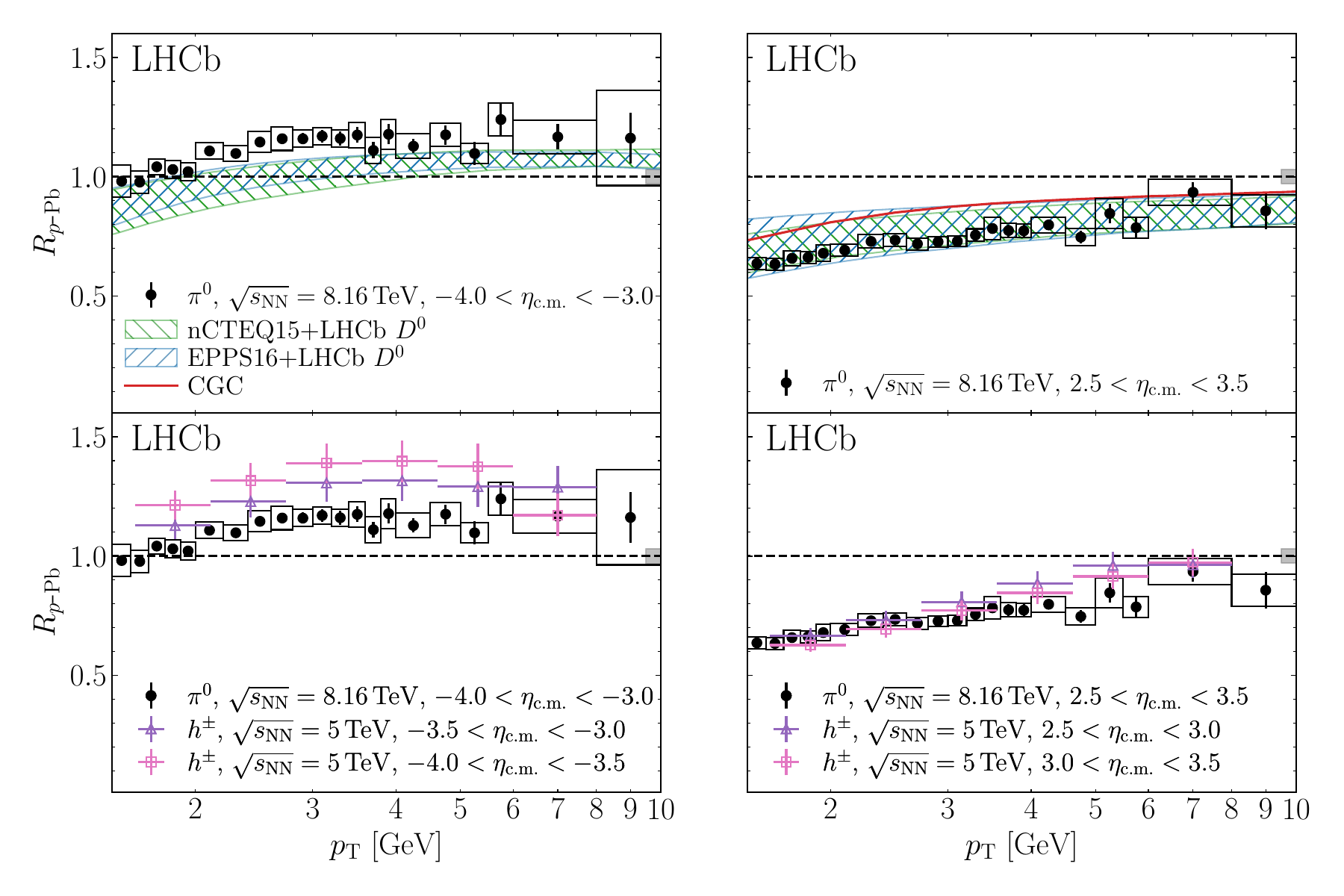}
     \caption{The $\pi^0$ nuclear modification factor measured by LHCb in $p{\rm
     Pb}$ collisions at $\sqrt{s_{\rm NN}}=8.16\,{\rm
     TeV}$~\cite{LHCb:2022vfn}.}
     \label{fig:pi0}
\end{figure}

Understanding this backward charged-particle enhancement is necessary in order
to understand the effects of low-$x$ phenomena on the forward charged-particle
data. The backward enhancement could be caused by a final-state effect, such as
radial flow. In this case, the production of different particle species could be
enhanced by different amounts~\cite{Pierog:2013ria}. As a result, studies of
identified particle production are necessary to untangle initial- and
final-state effects. The LHCb collaboration measured $\pi^0$ production in
$p{\rm Pb}$ collisions at $\sqrt{s_{\rm NN}}=8.16\,{\rm
TeV}$~\cite{LHCb:2022vfn}. The resulting nuclear modification factor is shown in
Figure~\ref{fig:pi0}. The forward data agree with both nPDF predictions and the
LHCb charged-particle data. The backward data, however, show a larger
enhancement than predicted by nPDF calculations, but a smaller enhancement than
that observed in the LHCb backward charged-particle data. This pattern is
qualitatively consistent with expectations from radial flow, and further studies
of identified particles will help clarify the source of the enhancement.

\section{Open charm production}

\begin{figure}[h!]
     \centering
     \includegraphics[width=\textwidth]{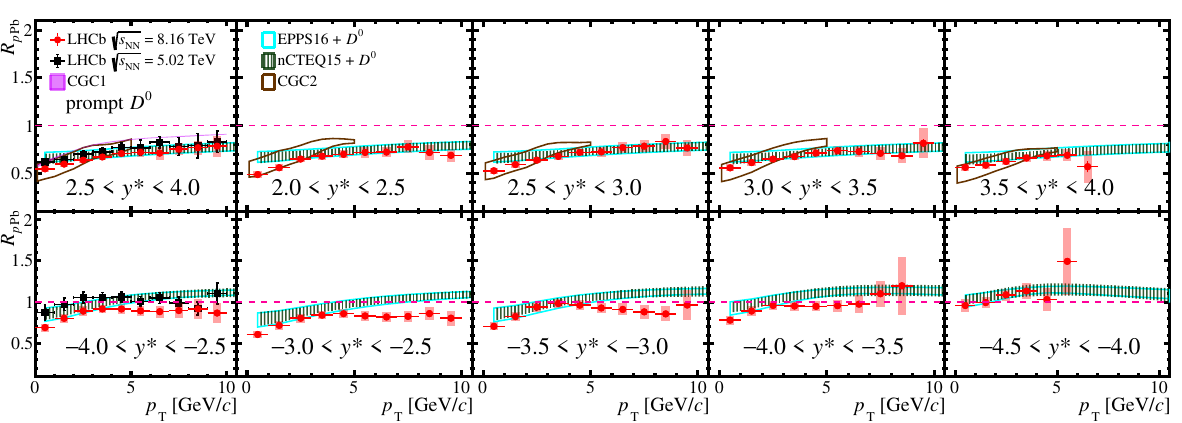}
     \caption{The $D^0$ nuclear modification factor measured by LHCb in $p{\rm
     Pb}$ collisions at $\sqrt{s_{\rm NN}}=8.16\,{\rm
     TeV}$~\cite{LHCb:2022rlh}.}
     \label{fig:d0}
\end{figure}

The LHCb collaboration recently measured $D^0$ production in $p{\rm Pb}$
collisions at $\sqrt{s_{\rm NN}}=8.16\,{\rm TeV}$~\cite{LHCb:2022rlh}. The
higher collision energy provides access to lower $x$ than the $5.02\,{\rm TeV}$
measurement. Furthermore, the $8.16\,{\rm TeV}$ dataset is much larger than the
$5.02\,{\rm TeV}$ dataset, allowing for much higher precision and granularity.
Results are shown in Figure~\ref{fig:d0}. The upper row of figures shows results
at forward rapidity, which agree well with both nPDF and CGC predictions. The
backward results, however, show tension with nPDF calculations at high $p_{\rm
T}$. This tension occurs in a kinematic region similar to the charged-particle
and $\pi^0$ enhancements.

\section{Ultraperipheral collisions}

\begin{figure}[h!]
     \centering
     \includegraphics[width=0.43\textwidth]{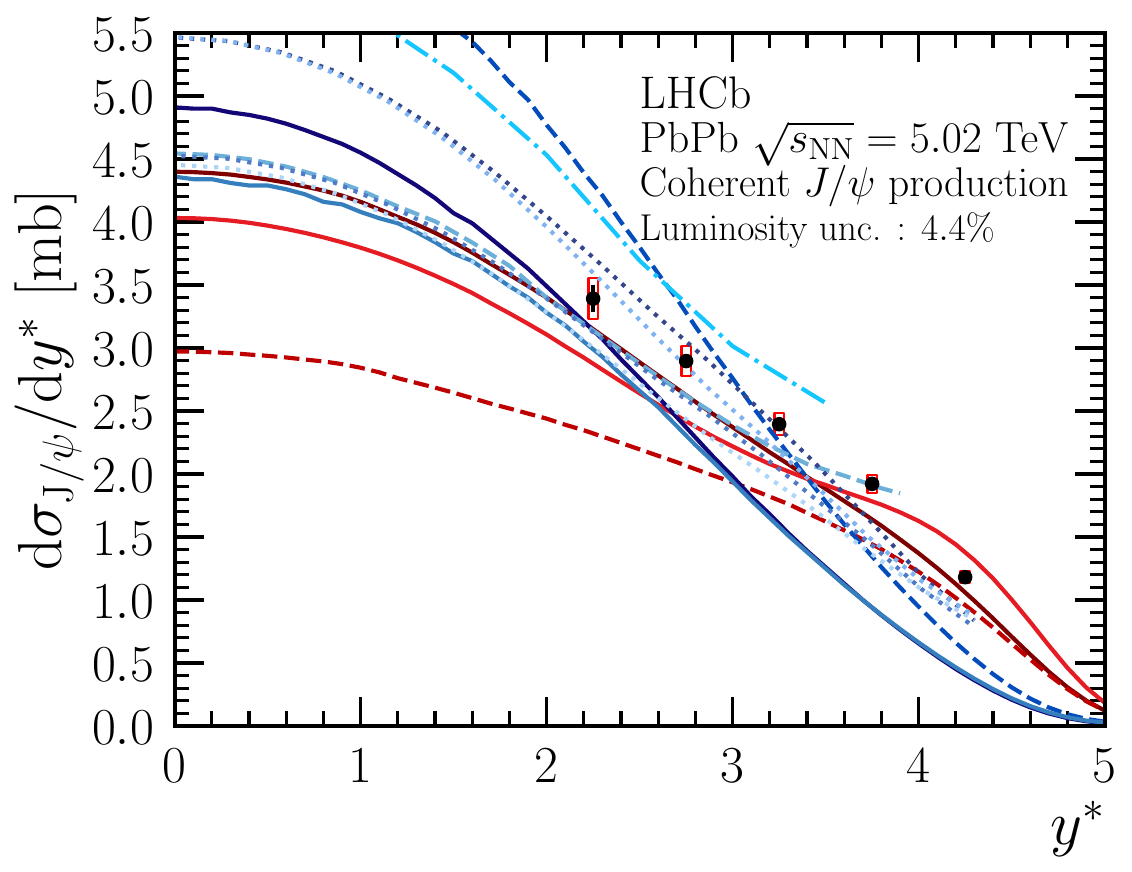}
     \includegraphics[width=0.56\textwidth]{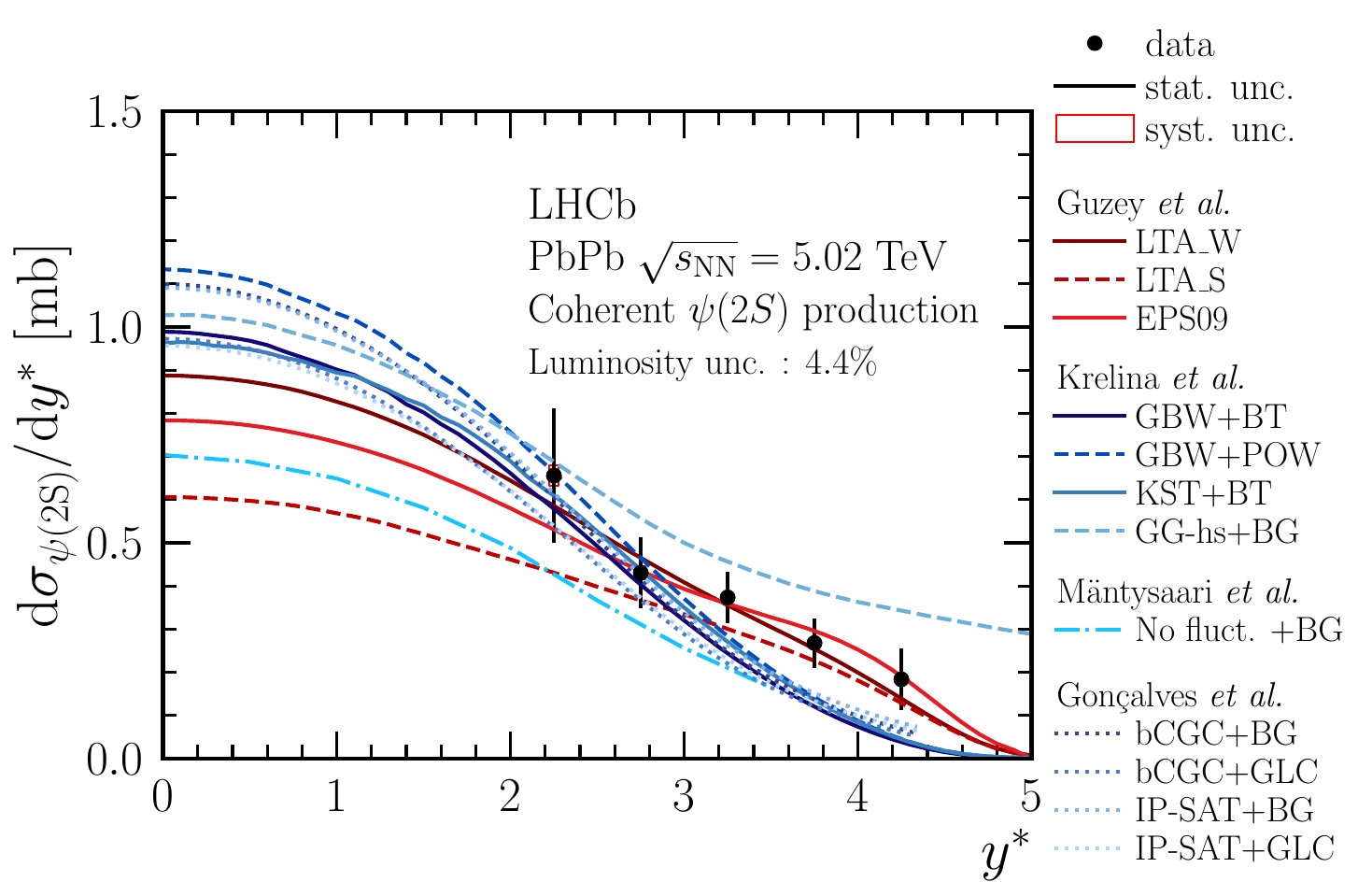}
     \caption{The (left) $J/\psi$ and (right) $\psi(2S)$ photoproduction cross
     sections measured in ${\rm PbPb}$ UPCs at $\sqrt{s_{\rm NN}}=5.02\,{\rm
     TeV}$~\cite{LHCb:2022ahs}.}
     \label{fig:upc}
\end{figure}

In addition to inclusive forward particle production in $p{\rm Pb}$ collisions,
the low-$x$ regime can also be probed using ultraperipheral heavy-ion collisions
(UPCs). In UPCs, a photon emitted by a nucleus interacts with the other
colliding nucleus. LHCb has studied vector meson production in UPCs in ${\rm
PbPb}$ collisions at $\sqrt{s_{\rm NN}}=5.02\,{\rm TeV}$\,\cite{LHCb:2022ahs}.
The $J/\psi$ and $\psi(2S)$ cross sections measured by the LHCb collaboration
are shown in Figure~\ref{fig:upc}. The vector meson photoproduction cross
section is proportional to the square of the gluon density in the nucleus at
leading order and probes $x\lesssim10^{-5}$~\cite{Jones:2015nna}. Perturbative
QCD calculations of UPC cross sections suffer from large scale
uncertainties~\cite{Eskola:2022vpi}, but the LHCb studies of $J/\psi$ and
$\psi(2S)$ production together constrain the scale dependence of low-$x$ physics
models.

\section{Conclusions}

LHCb data has allowed modern nPDFs to precisely describe the collinear structure
of the nucleon at $x\lesssim10^{-5}$. Despite this successful description, a
satisfactory understanding of the underlying low-$x$ physics remains elusive.
Even with newfound precision, perturbative QCD calculations relying on nPDFs and
linear QCD still often agree with CGC predictions. As a result, the task of
low-$x$ physics at the LHC is now to overconstrain parton densities across a
wide kinematic range in order to find the onset of nonlinear QCD effects.
Accomplishing this will require understanding final-state effects in
light-hadron production, which will benefit from future LHCb studies of
identified particle production in $p{\rm Pb}$ collisions. Measurements of light
neutral meson production will also pave the way for studies of low-$p_{\rm T}$
direct photon production at LHCb, which will directly probe the gluon nPDF at
low $x$ and $Q^2$. Finally, LHCb is planning to collect proton-oxygen data
during Run 3. These data will help determine the $A$ dependence of low-$x$
effects and could help reveal the onset of nonlinear QCD.

\section*{Acknowledgements}

This material is based on work supported by the U.S. National Science Foundation.

\end{document}